\documentclass[a4paper]{article}
\usepackage{amsmath,amssymb,amsthm}
\usepackage[utf8]{inputenc}
\usepackage[T1]{fontenc}
\usepackage{epsfig}
\usepackage[backend=biber,eprint=true,sorting=none]{biblatex}
\usepackage{hyperref}
\hypersetup{pdftitle={2209.14248.pdf}}

\newcommand{\wfb}{\bar{P}}
\newcommand{\myphi}{\varphi}
\newcommand{\mysharp}{\#}

\newcommand{\dummy}[1]{}
\newcommand{\zerointop}{\hat{U}_0}
\newcommand{\infintop}{\hat{U}_{\infty}}
\newcommand{\basissize}{W}
\newcommand{\slaterint}{y}

\DeclareMathOperator{\supp}{supp}
\DeclareMathOperator{\diag}{diag}

\begin{document}

\title{Electronic Structure Calculations with the Exact Pseudopotential %
  and Interpolating Wavelet Basis}

\author{Tommi H\"oyn\"al\"anmaa \\
  e-mail: tommi.hoynalanmaa@tuni.fi \\
  Niuvankuja 65 \\
  FI-70240, Kuopio, Finland \\
  \\
  Tapio T. Rantala \\
  e-mail: tapio.rantala@tuni.fi \\
  Physics, Tampere University, \\
  P. O. Box 692, FI-33101 Tampere, Finland}

\maketitle

\begin{abstract}
Electronic structure calculations are mostly carried out with Coulomb
potential singularity adapted basis sets like STO or contracted GTO.  With other
basis or for heavy elements the pseudopotentials may appear as a
practical alternative.  Here, we introduce the exact pseudopotential
(EPP) to remove the Coulomb singularity and test it for orbitals of
small atoms with the interpolating wave basis set.
  We apply EPP to the Galerkin method with a
  basis set consisting of Deslauriers--Dubuc scaling functions on the
  half-infinite real interval.  We demonstrate the EPP--Galerkin method
  by computing the hydrogen atom 1s, 2s, and 2p orbitals
  and helium atom configurations $\mathrm{He\;1s^2}$,
  $\mathrm{He\;1s2s\;{}^1 S}$, and $\mathrm{He\;1s2s\;{}^3 S}$.
  We compare the method to the ordinary interpolating wavelet Galerkin
  method (OIW--Galerkin)
  handling the singularity at the nucleus by excluding the scaling
  function located at the origin from the basis.
  We also compare the performance of our approach to that of finite--difference
  approach, which is another practical method for spherical atoms.
  We find the accuracy of the EPP--Galerkin method better than both of
  the above mentioned methods.
\end{abstract}


\section{Introduction}

The Coulomb singularity in the hamiltonian may appear as a challenge in electronic structure calculations.  Singularity adapted Slater type atomic orbitals (STO) basis is the usual solution to this, and also, gaussian type contracted basis functions (GTO) have turned out to be useful with sufficient accuracy.  The latter one is more popular due to other practical advantages.

Pseudopotentials removing the singularity are another type of solution to this problem.  In case the core electrons do not play an essential role in the problem at hand or valence electrons are expanded in plane waves, like it is with heavy elements or periodic crystalline systems.  In those cases the pseudopotentials typically replace the nuclei and a number of core electrons with their charge distribution, and possibly, some other core properties.

%When atoms and molecules are computed quantum physically the potential where the electrons move contains singularities at the nuclei. These singularities are usually handled by approximating the potential with a pseudopotential that removes those singularities.  Pseudopotentials are also used to approximate the total charge of the nuclei and core electrons so that it is sufficient to take only the valence electrons in the computations of the system.

One-dimensional interpolating wavelets have been used for atomic
computations for example in Ref.\ \cite{hrr2004}.
Fischer and Defranceschi \cite{fd1998} have also solved hydrogen-like atoms
with wavelets.
In Ref.\ \cite{hrr2004} we used
ordinary Deslauriers--Dubuc interpolating wavelets
\cite{dd1989,dubuc1986,cl1996,donoho1992interpolating,goedecker1998}
defined on the whole real axis so
including the negative real axis in the computations. We handled the
singularity at the origin by excluding the scaling
function at the origin from the basis.
We used the nonstandard operator form for the various operators needed
in the computations.
We computed the Schrödinger equation of hydrogenlike atoms (ions) and
Hartree--Fock equations of
some light many-electron atoms (helium, lithium, beryllium, neon,
sodium, magnesium, and argon). In this article we repeat similar
computations for hydrogen and helium atoms, but using a different
method to handle the singularity of the potential and only one
resolution level.
We handle the singularity by computing
the Schrödinger and Hartree--Fock equations for a range of variables
\(r \in \left[a,\infty\right[\) which does not contain the origin.
Here \(r\) is the position coordinate. The range
\(r \in \left[0,a\right]\) is neglected for hydrogen and for helium
its contribution to the Slater integrals is computed using the
hydrogenic orbitals.

Arias \cite{arias1999} and Engeness and Arias \cite{ea2002} developed
formalism for electronic structure calculations with interpolating
wavelets so that matrix elements of the operators are computed as
usual and overlap matrices are used in the matrix form of the
Schr\"odinger equation. On the other hand, we use the interpolating
dual scaling functions and wavelets for the computation of matrix
elements.

One-dimensional
interpolating multiresolution analysis in space
$C_\mathrm{u}(\mathbb{R})$ consisting of uniformly continuous bounded
functions in $\mathbb{R}$ has been constructed in Ref.\ \cite{cl1996}. 
One-dimensional
interpolating multiresolution analysis in space
$C_0(\mathbb{R})$ consisting of continuous functions in $\mathbb{R}$
vanishing at infinity has been constructed in \cite{donoho1992interpolating}.
Both of these constructions are based on Deslauriers-Dubuc functions
\cite{dd1989,dubuc1986}.
Donoho \cite{donoho1992interpolating} constructs wavelets on a finite real
interval, too.
We compute the eigenenergies of hydrogen atom 1s, 2s, and 2p orbitals
and helium atom configurations $\mathrm{He\;1s^2}$,
$\mathrm{He\;1s2s\;{}^1 S}$, and $\mathrm{He\;1s2s\;{}^3 S}$ with the
EPP method using both Galerkin method with interpolating
wavelets and finite difference method.

We denote the pointwise product of functions \(f\) and \(g\) by
\(f \star g\).
We use atomic units throughout this article 
(\(e = m_e = \hbar = 4 \pi \varepsilon_{0} = 1\))
and denote the atomic unit of energy by \(\mathrm{Ha}\) (Hartree).

%% We approximate the total charge of the core region \(r \leq r_0\) by
%% \(Z' = Z - z\) where \(Z\) is the charge of the nucleus and \(z\) is
%% the charge of the electrons in the core region.
%% We have
%% \begin{equation}
%%   z = \int_{r=0}^{r_0} \rho_{\mathrm{hydr}}(r) dr
%% \end{equation}
%% where \(\rho_{\mathrm{hydr}}\) is the hydrogenic charge density of the
%% electrons in the system.

\section{Interpolating Wavelets on Half-Infinite Interval}
\label{sec:basis}

\subsection{Interpolating Wavelets}

Interpolating wavelets are a biorthogonal wavelet family. Since the
dual scaling functions and dual wavelets of these functions are finite
sums of Dirac delta functions the matrix elements involving
interpolating wavelets usually require evaluating some function in a
finite set of points. An interpolating wavelet family is defined by a
mother scaling function $\myphi$, mother wavelet $\psi$, and four
finite filters $h_j$,
$g_j$, $\tilde{h}_j$, and $\tilde{g}_j$ where $j=-m,\ldots,m$. The
functions $\myphi$, $\psi$, $\tilde{\myphi}$, and $\tilde{\psi}$
satisfy equations
\begin{equation}
  \myphi(x) = \sum_{j=-m}^m h_j \myphi(2x-j) ,
\end{equation}
\begin{equation}
  \psi(x) = \sum_{j=-m}^m g_j \myphi(2x-j) ,
\end{equation}
\begin{equation}
  \tilde{\myphi}(x) = \sum_{j=-m}^m \tilde{h}_j \tilde{\myphi}(2x-j) ,
\end{equation}
and
\begin{equation}
  \tilde{\psi}(x) = \sum_{j=-m}^m \tilde{g}_j \tilde{\myphi}(2x-j) .
\end{equation}
The two-index basis functions and dual basis functions are
\begin{equation}
  \myphi_{j,k}(x) = \myphi(2^j x - k) ,
\end{equation}
\begin{equation}
  \psi_{j,k}(x) = \psi(2^j x - k) ,
\end{equation}
\begin{equation}
  \tilde{\myphi}_{j,k}(x) = 2^j \tilde{\myphi}(2^j x - k) ,
\end{equation}
and
\begin{equation}
  \tilde{\psi}_{j,k}(x) = 2^j \tilde{\psi}(2^j x - k) .
\end{equation}
A wavelet basis consists of scaling functions \(\myphi_{j_0,k}\),
\(k \in \mathbb{Z}\), and wavelets \(\psi_{j,k}\), \(j \geq j_0\), \(k
\in \mathbb{Z}\), where \(j_0 \in \mathbb{Z}\) is the minimum
resolution level. The expansion of an arbitrary (regular enough)
function \(f : \mathbb{R} \to \mathbb{R}\) in the wavelet basis is
\begin{equation}
  f(x) = \sum_{k \in \mathbb{Z}} s_k \myphi_{j_0,k}(x) +
  \sum_{j \geq j_0} \sum_{k \in \mathbb{Z}} d_{j,k} \psi_{j,k}(x) .
\end{equation}

\subsection{The Basis Set}

This derivation is based on section 3 in \cite{donoho1992interpolating}. We
construct a basis set on half-infinite interval \(\mathbb{R}_0 = \{ r
\geq 0 \vert r \in \mathbb{R} \}\).
We define \(\myphi\) to be a Deslauriers-Dubuc scaling function of
some order \(D\) and \(\myphi_{j,k}(x) := \myphi(2^j x - k)\) for
\(j, k \in \mathbb{Z}\).
We define a wavelet expansion of a function \(f\) on \(\mathbb{R}_0\) by
\begin{equation}
  \label{eq:expansion-1}
  \tilde{f} := \sum_{k = 0}^D \beta_{j,k} \myphi^{\mysharp}_{j,k}
  + \sum_{k=D+1}^{\infty} \beta_{j,k} \myphi_{j,k} .
\end{equation}
When we use a finite basis of size \(\basissize\) we have
\begin{equation}
  \label{eq:expansion-2}
  \tilde{f} := \sum_{k = 0}^D \beta_{j,k} \myphi^{\mysharp}_{j,k}
  + \sum_{k=D+1}^{\basissize-1} \beta_{j,k} \myphi_{j,k}
\end{equation}
We must have \(\basissize > 2D\) so that functions
\(\myphi^{\mysharp}_{j,k}(x)\), \(0 \leq k \leq D\), vanish for
\(x \geq 2^{-j} \basissize\). This kind of truncation of the basis
requires that the function \(f\) approximately vanishes for
\(x > 2^{-j} W\).

Suppose that we are given samples \(\beta_{j,k} = f(2^{-j} k)\) for
\(k \in \mathbb{N}\) and \(f\) is some function from \([0,\infty[\)
into \(\mathbb{R}\).
We define \(\pi^\mysharp_j\) to be the polynomial of degree \(D\) for
which \(\pi^\mysharp_j(2^{-j}k) = f(2^{-j}k)\) for all
\(k = 0,\ldots,D\).
We define
\begin{equation}
  \tilde{\beta}_{j,k} := \pi^{\mysharp}_j(2^{-j} k)
\end{equation}
for \(k < 0\) and
\begin{equation}
  \tilde{\beta}_{j,k} := \beta_{j,k}
\end{equation}
for \(k \geq 0\).
Now \(f\) can be extrapolated onto the whole real line by
\begin{equation}
  \tilde{f} = \sum_{k=-\infty}^\infty \tilde{\beta}_{j,k} \myphi_{j,k} .
\end{equation}

As each coefficient \(\tilde{\beta}_{j,k}\) is a linear functional of
coefficients \(\beta_{j,k'}\) we may define extrapolation weights
\(e^{\mysharp}_{k,k'}\) so that
\begin{equation}
  \tilde{\beta}_{j,k} = \sum_{k'=0}^D e^{\mysharp}_{k,k'} \beta_{j,k'}
\end{equation}
for \(k < 0\).
When \(f := \myphi_{j,l}\) we have
\begin{equation}
  \tilde{\beta}_{j,k} = e^{\mysharp}_{k,l}
\end{equation}
where \(l \in \{ 0, \ldots, D \}\).
Consequently the quantities \(e^{\mysharp}_{k,l}\) can be computed by
polynomial interpolation of functions \(\myphi_{j,l}\). As
\begin{equation}
  \supp \myphi_{j,k} \subset 2^{-j} [ k - D, k + D ]
\end{equation}
we need only values \(k \in \{ -D, \ldots, -1 \} \).
We define
\begin{equation}
  \myphi^{\mysharp}_{j,k} := \myphi_{j,k} + \sum_{l<0} e^{\mysharp}_{l,k} \myphi_{j,l}
\end{equation}
for \(k = 0,\ldots,D\).
Note that
\begin{equation}
  \left\langle \tilde{\myphi}_{j,k} , \myphi^{\mysharp}_{j,l}
  \right\rangle = \delta_{k,l}
\end{equation}
for \(k \geq 0\) and \(0 \leq l \leq D\).
Let \(A\) be a linear operator from \(C_0(\mathbb{R})\) to
\(C_0(\mathbb{R})\). The matrix elements \(A_{k,l}\), \(l=0,\ldots,D\)
are given by
\begin{equation}
  \left\langle \tilde{\myphi}_{j,k}, A \myphi^{\mysharp}_{j,l} \right\rangle =
  \left\langle \tilde{\myphi}_{j,k}, A \myphi_{j,l} \right\rangle
  +
  \sum_{\alpha < 0} e^{\mysharp}_{\alpha,l}
  \left\langle \tilde{\myphi}_{j,k}, A \myphi_{j,\alpha} \right\rangle .
\end{equation}
Let \(v(f)\) denote the coefficient vector
\((\beta_{j,k})_{k=0}^{W-1}\) defined by equation
\eqref{eq:expansion-2} and define
\begin{equation}
  M(f) := \left(f(2^{-j} k) \delta_{k,k'}\right)_{k,k'=0}^{W-1,W-1}
\end{equation}
for some function \(f : \mathbb{R} \to \mathbb{R}\).

\section{Schrödinger Equations of Hydrogen-like Atoms and Helium Atom
  in the EPP-Wavelet Basis}

\subsection{General}

Suppose that we have a system consisting of a positively charged
nucleus at the origin and \(N\) electrons.
In EPP method we choose some small radius \(r_0\) so that inside the
sphere with radius \(r_0\) the wavefunctions of the system are approximated by
hydrogenic wavefunctions and the actual computations are done only for
values \(r \geq r_0\). Actually we define a basis set for
half-infinite interval \(\left[0, \infty\right[\) and make a change of
variables \(s = r - r_0\). For Hartree--Fock calculations the Slater
integrals are computed by
\begin{eqnarray}
  \label{eq:slater}
  \nonumber
  \bar{\slaterint}_{\mathrm{ab}}^0(s) & = & \frac{Q_{\mathrm{ab}}}{s+r_0} +
  \frac{1}{s+r_0} \int_0^s \bar{P}_{\mathrm{a}}(s')
  \bar{P}_{\mathrm{b}}(s') ds' \\
  & & + \int_s^\infty \frac{1}{s'+r_0} \bar{P}_{\mathrm{a}}(s')
  \bar{P}_{\mathrm{b}}(s')
  ds'
\end{eqnarray}
where \(s \geq 0\) and \(Q_{\mathrm{ab}}\) is a system-dependent
quantity that approximates the contribution of the EPP core region to
the Slater integral.

\subsection{Hydrogen-like Atoms}

The Schrödinger equation of
the hydrogen atom and Hartree--Fock equations of atoms
\cite{saad2010numerical,af2005,schmidt97,cowan81}
and representation in the interpolating wavelet basis \cite{hrr2004}
is our starting point.
With a change of variables \(s := r - r_0\) the Schrödinger equation
of a hydrogen-like atom in interval \(r \geq r_0\) takes the
form
\begin{equation}
  \label{eq:schrodinger}
  \left(
  - \frac{1}{2} \frac{d^2}{ds^2} - \frac{Z}{s + r_0} +
  \frac{l(l+1)}{2^(s + r_0)^2}
  \right)
  \wfb(s) = E \wfb(s), \;\;\;\; s \geq 0
\end{equation}
where \(Z\) is the charge of the nucleus, \(l\) is the angular
momentum quantum number, and
\begin{math}
  \wfb(s) = P(r_0 + s)
\end{math}
for \(s \geq 0\).

We define the second derivative filter by
\begin{equation}
  a_k := \left\langle \tilde{\myphi} , D^2 \myphi(\cdot - k) \right\rangle
\end{equation}
Matrix elements of the Laplacian operator \(L\) are computed by
\begin{equation}
  L_{k,l} := \left\langle \tilde{\myphi}_{j,k} , L
  \myphi^{\mysharp}_{j,l} \right\rangle
  = 2^{2j} \left( a_{l-k} + \sum_{\alpha=-D}^{-1}
  e^{\mysharp}_{\alpha,l} a_{\alpha - k} \right)
\end{equation}
for \(0 \leq l \leq D\) and
\begin{equation}
  L_{k,l} := \left\langle \tilde{\myphi}_{j,k} , L
  \myphi_{j,l} \right\rangle
  = 2^{2j} a_{l-k}
\end{equation}
for \(D < l < W\). Note that matrix \(L\) is generally not hermitian.
The potential energy operator is computed as a diagonal matrix
\begin{equation}
  \hat{V}_{k,k} = V(2^{-j}k)
\end{equation}
where
\begin{equation}
  V(y) = - \frac{Z}{y + a}
\end{equation}
for \(y \geq 0 \).
The centrifugal potential is computed in the same way.

\subsection{Hartree--Fock Equations for Helium Atom}

Define the Slater integrals as
\begin{equation}
  \slaterint^0_{\mathrm{ab}}(r) = \int_{r'=0}^\infty P_{\mathrm{a}}(r') \gamma(r,r') P_{\mathrm{b}}(r') dr'
\end{equation}
where a and b denote the atomic orbitals and
\begin{equation}
  \gamma(r,r') = \frac{1}{\max\{r,r'\}} .
\end{equation}
We use symbol \(y\) instead of \(Y\) to avoid confusion with spherical
harmonics.
By doing a similar change of variables \(s := r - r_0\) the
Hartree--Fock equation of the ground state of the helium atom in
interval \(r \geq r_0\) takes the form
\begin{equation}
  \label{eq:he-ground-state}
  \left(
  - \frac{1}{2} \frac{d^2}{ds^2} - \frac{2}{s + r_0}
  + \slaterint^0_{\mathrm{1s1s}}(s+r_0)
  \right)
  \wfb_{\mathrm{1s}}(s) = \varepsilon_{\mathrm{1s}}
  \wfb_{\mathrm{1s}}(s), \;\;\;\; s \geq 0 .
\end{equation}
The Hartree--Fock equations for the helium 1s2s ${}^1\mathrm{S}$
configuration are
\begin{eqnarray}
  \label{eq:he1s2s1S-1s}
  \left( -\frac{1}{2} \frac{d^2}{ds^2} - \frac{2}{s + r_0} +
  \slaterint^0_{\mathrm{2s2s}}(s+r_0) \right) \bar{P}_{\mathrm{1s}}(s) =
  \varepsilon_{\mathrm{1s}} \bar{P}_{\mathrm{1s}}(s)
  \\
  \label{eq:he1s2s1S-2s}
  \left( -\frac{1}{2} \frac{d^2}{ds^2} - \frac{2}{s + r_0} +
  \slaterint^0_{\mathrm{1s1s}}(s+r_0) \right) \bar{P}_{\mathrm{2s}}(s) =
  \varepsilon_{\mathrm{2s}} \bar{P}_{\mathrm{2s}}(s)
\end{eqnarray}
and for helium 1s2s ${}^3\mathrm{S}$ configuration
\begin{eqnarray}
  \nonumber
  \left( -\frac{1}{2} \frac{d^2}{ds^2} - \frac{2}{s + r_0} +
  \slaterint^0_{\mathrm{2s2s}}(s+r_0) \right) \bar{P}_{\mathrm{1s}}(s) & = &
  \varepsilon_{\mathrm{1s}} \bar{P}_{\mathrm{1s}}(s)
  + \\
  \label{eq:he1s2s3S-1s}
  & & \slaterint^0_{\mathrm{1s2s}}(s+r_0) \bar{P}_{\mathrm{2s}}(s)
  \\
  \nonumber
  \left( -\frac{1}{2} \frac{d^2}{ds^2} - \frac{2}{s + r_0} +
  \slaterint^0_{\mathrm{1s1s}}(s+r_0) \right) \bar{P}_{\mathrm{2s}}(s) & = &
  \varepsilon_{\mathrm{2s}} \bar{P}_{\mathrm{2s}}(s)
  + \\
  \label{eq:he1s2s3S-2s}
  & & \slaterint^0_{\mathrm{2s1s}}(s+r_0) \bar{P}_{\mathrm{1s}}(s)
\end{eqnarray}

\subsection{EPP of Helium Atom}

We define \(P_{\mathrm{a}}(r)\) to be the exact Hartree--Fock
wavefunction of the orbital \(a\) of the atom.
We define the operators \(\zerointop\) and \(\infintop\)
\cite{hrr2004} by
\begin{equation}
  (\zerointop f)(s) = \int_0^s f(s') ds'
\end{equation}
and
\begin{equation}
  (\infintop f)(s) = \int_s^{\infty} f(s') ds'
\end{equation}
Define \(P_{\mathrm{1s,H}}(r')\) and \(P_{\mathrm{2s,H}}(r')\)
to be the hydrogenic orbitals of the helium atom.
Then we have
\begin{equation}
  \langle \tilde{\myphi}_{j,k} \zerointop \myphi_{j,l} \rangle
  = 2^{-j} (\Phi(\vert k \vert -l)-\Phi{-l})
\end{equation}
and
\begin{equation}
  \langle \tilde{\myphi}_{j,k} \infintop \myphi_{j,l} \rangle
  = 2^{-j} (1 - \Phi(\vert k \vert -l))
\end{equation}
where
\begin{equation}
  \Phi(x) = \int_{-\infty}^x \myphi(y) dy .
\end{equation}
The Slater integrals in the shifted variables are obtained from
equation \eqref{eq:slater} where we set
\begin{equation}
  Q_{\mathrm{ab}} :=
  \frac{\bar{P}_{\mathrm{a}}(0)}{P_{\mathrm{a,H}}(r_0)}
  \frac{\bar{P}_{\mathrm{b}}(0)}{P_{\mathrm{b,H}}(r_0)}
  \int_0^{r_0} P_{\mathrm{a,H}}(r') P_{\mathrm{b,H}}(r') dr' .
\end{equation}
for the helium ground state,
and
\begin{equation}
  Q_{\mathrm{ab}} := \int_0^{r_0} P_{\mathrm{a,H}}(r') P_{\mathrm{b,H}}(r') dr' .
\end{equation}
for the excited states of helium.
Define
\begin{equation}
  q(s) := \frac{1}{s+r_0}
\end{equation}
and
\begin{equation}
  S_0 := M(q) U_0 + U_\infty M(q) .
\end{equation}
Now
\begin{equation}
  v(\bar{\slaterint}^0_{\mathrm{ab}}) =
  Q_{\mathrm{ab}} v(q) + S_0 \left(v(\bar{P}_{\mathrm{a}} \star
  \bar{P}_{\mathrm{b}})\right)
\end{equation}
where \(U_0\) and \(U_\infty\) are the matrices of operators
\(\zerointop\) and \(\infintop\) in the basis set constructed in
section \ref{sec:basis}.
We define \(\mathbf{v}_{\mathrm{a}} = v(\bar{P}_{\mathrm{a}})\) and
\(\mathbf{v}_{\mathrm{b}} = v(\bar{P}_{\mathrm{b}})\).
The matrix of the exchange integral operator
\begin{equation}
  (\hat{K}_{\mathrm{a}} \bar{P}_{\mathrm{a}})(s) :=
  \bar{\slaterint}^0_{\mathrm{ab}}(s) \bar{P}_{\mathrm{b}}(s)
\end{equation}
is computed by
\begin{equation}
  K_{\mathrm{a}} :=
  W_{\mathrm{a}} + M(\mathbf{v}_b) S_0 M(\mathbf{v}_b)
\end{equation}
The term $W_{\mathrm{a}} \mathbf{v}_a$ approximates the first term in
equation \eqref{eq:slater} as a linear function of $\mathbf{v}_a$.
In order to do this we approximate the wavefunction
\(P_{\mathrm{a}}(r)\) in region
\(r \in [0,r_0]\) by a linear function that is zero at the origin and
$\bar{P}_{\mathrm{a}}(0)$ at $r_0$. We have
\begin{equation}
  \frac{Q_{\mathrm{ab}}}{s + r_0} \bar{P}_\mathrm{b}(s) \approx
  \frac{1}{s + r_0} \bar{P}_{\mathrm{a}}(0) \left( \int_{s'=-r_0}^0 \left(1 +
  \frac{s'}{r_0}\right) P_{\mathrm{b,H}}(s'+r_0) ds' \right)
  \bar{P}_{\mathrm{b}}(s)
\end{equation}
The wavefunction $\bar{P}_{\mathrm{b}}(s)$ is taken from the previous
step of the Hartree--Fock iteration.
By approximating the wavefunctions by hydrogenic ones we get the
hydrogenic Slater integrals
\begin{eqnarray}
  \slaterint^0_{\mathrm{1s1s,H}}(r) & = & \frac{1}{r} - e^{-2Zr} \left(
  \frac{1}{r} + Z \right) \\
  \slaterint^0_{\mathrm{2s2s,H}}(r) & = &
  \frac{1}{r} + e^{-Zr} \left(
  -\frac{Z^3}{8} r^2 - \frac{1}{4} Z^2 r - \frac{3Z}{4} - \frac{1}{r}
  \right) \\
  \slaterint^0_{\mathrm{1s2s,H}}(r) & = &
  \frac{1}{27 \sqrt{2}} \left( 12 Z^2 r + 8Z \right) e^{-3Zr/2}
\end{eqnarray}
for \(r \geq 0\).
The scalar products involving the Slater integrals are approximated as
\begin{eqnarray}
  \nonumber
  \left\langle P_{\mathrm{a}} \vert \slaterint^0_{\mathrm{ab}}
  \vert P_{\mathrm{b}} \right\rangle
  & \approx &
  \left(\frac{\bar{P}_{\mathrm{a}}(0)}{P_{\mathrm{a,H}}(r_0)}\right)^2
  \left(\frac{\bar{P}_{\mathrm{b}}(0)}{P_{\mathrm{b,H}}(r_0)}\right)^2 \\
  & &
  \nonumber
  \cdot
  \int_{r'=0}^{r_0} P_{\mathrm{a,H}}(r')
  \slaterint^0_{\mathrm{ab,H}}(r') P_{\mathrm{b,H}}(r') dr' \\
  & &
  + \int_{s=0}^\infty \bar{P}_{\mathrm{a}}(s)
  \bar{\slaterint}^0_{\mathrm{ab}}(s) \bar{P}_{\mathrm{b}}(s) ds
\end{eqnarray}
for the helium ground state and
\begin{eqnarray}
  \nonumber
  \left\langle P_{\mathrm{a}} \vert \slaterint^0_{\mathrm{ab}}
  \vert P_{\mathrm{b}} \right\rangle
  & \approx &
  \int_{r'=0}^{r_0} P_{\mathrm{a,H}}(r')
  \slaterint^0_{\mathrm{ab,H}}(r') P_{\mathrm{b,H}}(r') dr' \\
  & &
  + \int_{s=0}^\infty \bar{P}_{\mathrm{a}}(s)
  \bar{\slaterint}^0_{\mathrm{ab}}(s) \bar{P}_{\mathrm{b}}(s) ds .  
\end{eqnarray}
for the excited states of helium.
%%We have \(E_{\mathrm{a,H}} = Z^2/(2n^2)\) where \(n\) is
%%the principal quantum number of the orbital \(a\).

\subsection{Total Energy of Helium Atom}

The total energy of the ground state of the helium atom is
\begin{equation}
  E(\mathrm{He\;1s^2}) = 2 \varepsilon_{\mathrm{1s}}
  -
  \left\langle P_{\mathrm{1s}} \vert \slaterint^0_{\mathrm{1s1s}}
  \vert P_{\mathrm{1s}} \right\rangle
\end{equation}
The total energy of the $\mathrm{1s2s\;{}^1 S}$ configuration of the
helium atom is
\begin{eqnarray}
  \nonumber
  E(\mathrm{He\;1s2s\;{}^1 S}) & = &
  \varepsilon_{\mathrm{1s}}
  + \varepsilon_{\mathrm{2s}} \\
  & & -\frac{1}{2}
  \left\langle P_{\mathrm{1s}} \vert \slaterint^0_{\mathrm{2s2s}} \vert
  P_{\mathrm{1s}} \right\rangle
  -\frac{1}{2}
  \left\langle P_{\mathrm{2s}} \vert \slaterint^0_{\mathrm{1s1s}} \vert
  P_{\mathrm{2s}} \right\rangle  
\end{eqnarray}
and for the $\mathrm{1s2s\;{}^3 S}$ configuration
\begin{eqnarray}
  \nonumber
  E(\mathrm{He\;1s2s\;{}^3 S}) & = &
  \varepsilon_{\mathrm{1s}}
  + \varepsilon_{\mathrm{2s}} \\
  \nonumber
  & &
  -\frac{1}{2}
  \left\langle P_{\mathrm{1s}} \vert \slaterint^0_{\mathrm{2s2s}} \vert
  P_{\mathrm{1s}} \right\rangle
  -\frac{1}{2}
  \left\langle P_{\mathrm{2s}} \vert \slaterint^0_{\mathrm{1s1s}} \vert
  P_{\mathrm{2s}} \right\rangle \\
  & &
  + \left\langle P_{\mathrm{1s}} \vert \slaterint^0_{\mathrm{1s2s}} \vert
  P_{\mathrm{2s}} \right\rangle  
  .
\end{eqnarray}

\section{Combination of EPP with Finite Difference Method}

The Schrödinger and Hartree--Fock equations are converted to matrix
equations using the biorthogonality relations of interpolating
wavelets \cite{hrr2004}. We compare these computations with the
Finite Difference Method, which is a straightforward method for solving
differential equations. The spatial and time domains are discretized
and derivative at a point is computed with a stencil applied to the
nearby points. This way the differential equation is converted to a
matrix equation. The Laplacian operator is approximated by
\begin{equation}
  u''(x) \approx \frac{u(x-h) - 2 u(x) + u(x+h)}{h^2}
\end{equation}
where $h$ is the discretization step size.

We discretize the Schrödinger equation \eqref{eq:schrodinger} at
points
\begin{math}
  p_j = jh
\end{math},
\begin{math}
  j = 0, \ldots, J + 1
\end{math}
where \(J\) is the number of actual computation points and \(h \in
\mathbb{R}_+\) is the grid spacing. We define the discretized
potential by \(v_j = \bar{V}(s_j)\).
The boundary condition at the end
of the interval is set by \(p_{J+1} = 0\). We have
\begin{equation}
  - \frac{p_{j+1} + (-2 - 2 h^2 v_j) p_j + p_{j-1}}{2h^2} = E p_j
\end{equation}
for \(j = 2, \ldots, J\).
We handle case \(j = 1\) by extrapolating \(p_0\) linearly from
\(p_1\) and \(p_2\). We get \(p_0 = 2 p_1 - p_2\) from which it
follows that
\begin{math}
  p_2 + (-2 - 2 h^2 v_j) p_1 + p_0 = -2 h^2 v_1 p_1
\end{math}.
Hence the difference equation for \(j = 1\) is
\begin{equation}
  v_1 p_1 = E p_1 .
\end{equation}

In order to discretize the exchange operator $\hat{K}_a$ we need to
discretize the integral operators
\begin{equation}
  (\hat{I}_g(f))(s) = \int_0^s g(s') f(s') ds' 
\end{equation}
and
\begin{equation}
  (\hat{I}_g^{\textrm{compl}}(f))(s) = \int_s^\infty g(s') f(s') ds' .
\end{equation}
We define
\begin{equation}
  (I(g))_{j,k} :=
  \left\{
  \begin{array}{ll}
    h g_k ; \;\;\;\; & k < j \\
    0 ; \;\;\;\; & k \geq j
  \end{array}
  \right.
\end{equation}
and
\begin{equation}
  (I^{\textrm{compl}}(g))_{j,k} :=
  \left\{
  \begin{array}{ll}
    h g_k ; \;\;\;\; & k \geq j \\
    0 ; \;\;\;\; & k < j
  \end{array}
  \right.
\end{equation}
where \(g_k = g(s_k)\). When \(f\) is a real function define
\(w(f) := \left(f(s_k)\right)_{k=1}^J\). Now the matrix of the
exchange integral operator is computed by
\begin{equation}
  K_a := W_a + K_a^0
\end{equation}
where \(W_a\) is computed as in the case of wavelets,
\begin{equation}
  K_a^0 := \diag(w(f_1)) I(\bar{P}_b)
  + I^{\textrm{compl}}(f_1 \star \bar{P}_b) ,
\end{equation}
and
\begin{equation}
  f_1(s) := \frac{1}{s + r_0} , \;\;\;\; s \geq 0 .
\end{equation}

\section{Demonstration and Test Results}

We demonstrate the EPP method by doing computations where the EPP
radius \(r_0\) and the basis size \(\basissize\) are varied.
We actually select a length scale \(u = R / W\) and
do a change of variables \(s = us'\) in equations
\eqref{eq:schrodinger}, \eqref{eq:he-ground-state},
\eqref{eq:he1s2s1S-1s}, \eqref{eq:he1s2s1S-2s},
\eqref{eq:he1s2s3S-1s}, and \eqref{eq:he1s2s3S-2s}.
The length scale \(u\) specifies how many atomic units of length a
length unit in our own coordinate system is.
Here \(R\) is the size of the computation domain.
For hydrogen 1s we have \(R = 15 \;\mathrm{a.u.}\), for hydrogen 2s
and 2p \(R = 25 \;\mathrm{a.u.}\), for He \(\mathrm{1s}^2\)
\(R = 15 \;\mathrm{a.u.}\), and for He 1s2s \({}^1\mathrm{S}\) and
\({}^3\mathrm{S}\) \(R = 20 \;\mathrm{a.u.}\).  We also set \(j = 0\)
for the basis set (see section \ref{sec:basis}).  The relative errors
of the quantities are given as
\begin{equation}
  \varepsilon = \left\vert \frac{x_{\mathrm{computed}} -
    x_{\mathrm{exact}}}{x_{\mathrm{exact}}} \right\vert .
\end{equation}
The amount of discontinuity of a computed wavefunction at point \(r =
r_0\) is measured by computing the relative error of the computed
wavefunction value \(\bar{P}(0)\) compared to the hydrogenic
wavefunction value \(P_{\mathrm{H}}(r_0)\).

%% \begin{table}
%%   \centering
%%   \begin{tabular}{lr}
%%     System & Energy / Ha \\
%%     \hline
%%     H 1s & -0.5 \\
%%     H 2s & -0.125 \\
%%     H 2p & -0.125 \\
%%     He $\mathrm{1s^2}$ & -2.8616800 \cite{ff1977} \\
%%     %% He $\mathrm{1s2s\;{}^1S}$ & -2.147 \cite{he1s2s} \\
%%     %% He $\mathrm{1s2s\;{}^3S}$ & -2.171 \cite{he1s2s}
%%     He $\mathrm{1s2s\;{}^1S}$ & -2.153 \\
%%     He $\mathrm{1s2s\;{}^3S}$ & -2.174
%%   \end{tabular}
%%   \caption{The exact energies and HF limits of H and He.
%%     Exact analytical results are given for hydrogenic orbitals and
%%     HF limits for other systems.
%%   }
%%   \label{tab:exact-results}
%% \end{table}

\begin{figure}
%%  \centering \includegraphics[scale=0.5]{figure3_1.eps}
  \centering \includegraphics{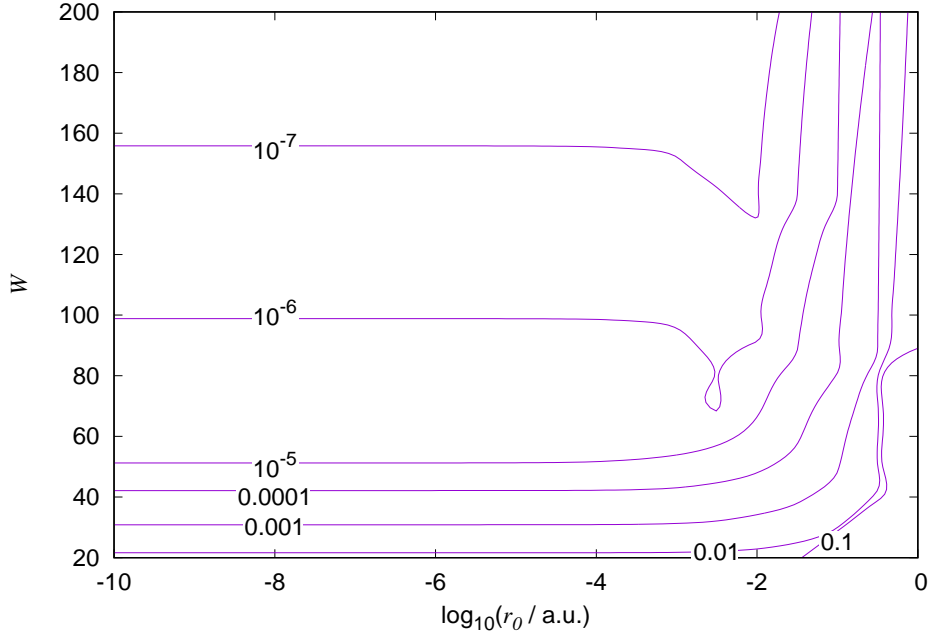}
  \caption{Hydrogen 1s orbital eigenenergy relative error.  The
    $r_0$ is the EPP radius in atomic units and $W$ is the
    basis size.  }
  \label{fig:1}
\end{figure}

\begin{figure}
%%  \centering \includegraphics[scale=0.5]{figure3_2.eps}
  \centering \includegraphics{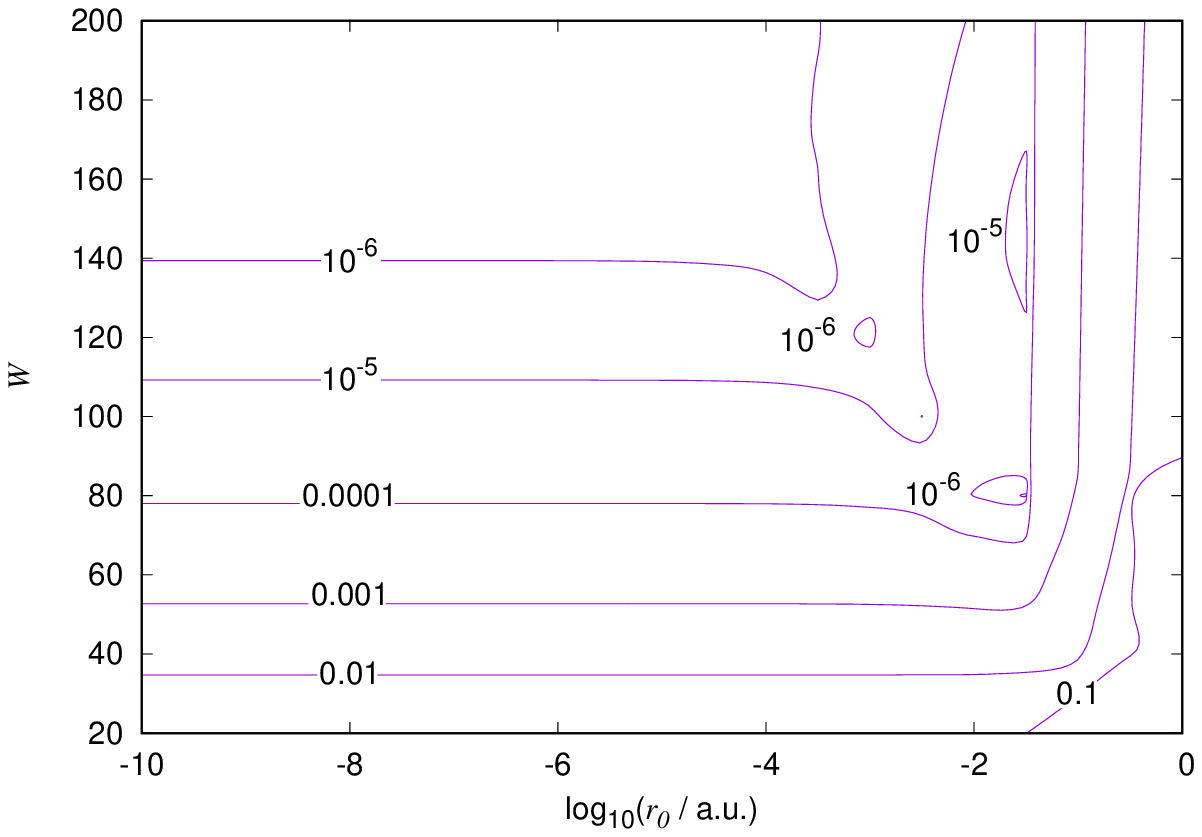}
  \caption{Relative error of the wavefunction value at the core radius
    for the hydrogen 1s orbital.  Notations as in Figure~\ref{fig:1}.
  }
  \label{fig:2}
\end{figure}

\begin{figure}
  \centering
  \includegraphics{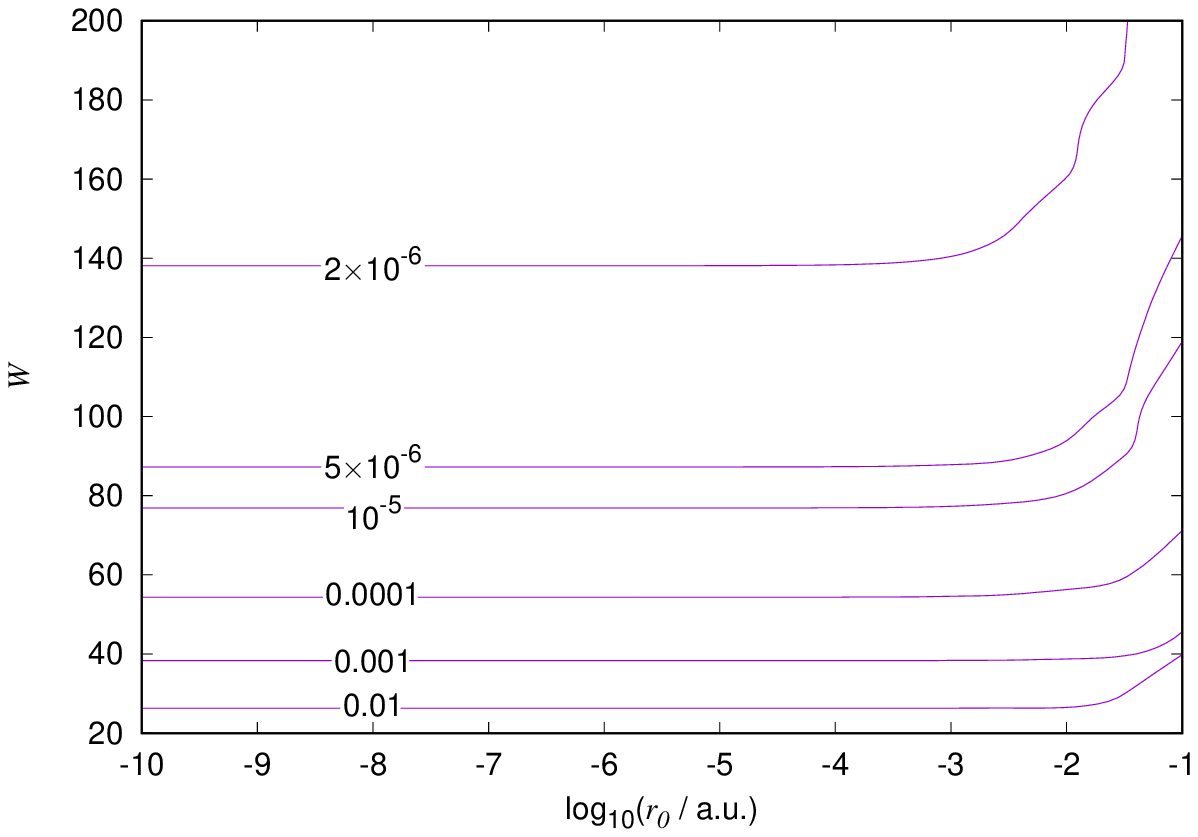}
  \caption{Hydrogen 2s orbital eigenenergy relative error.
    Notations as in Figure~\ref{fig:1}.
  }
  \label{fig:3}
\end{figure}

\begin{figure}
  \centering
  \includegraphics{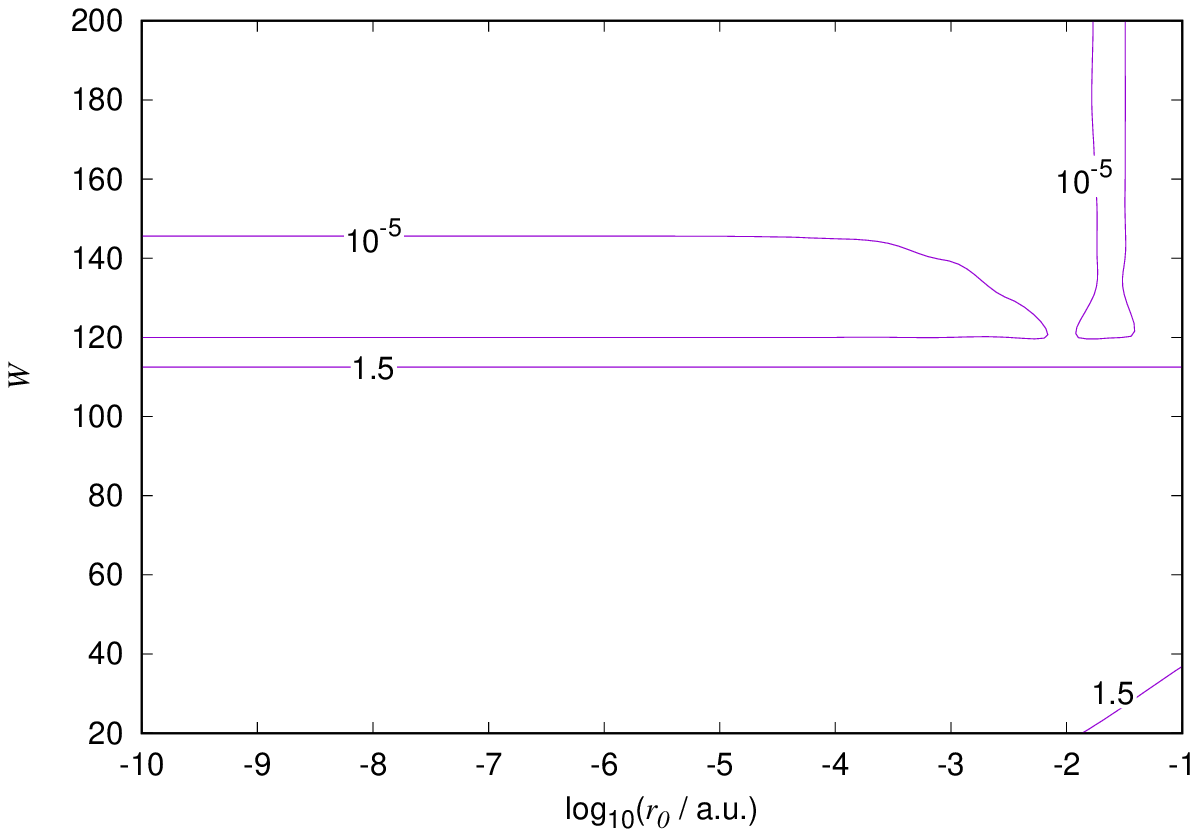}
  \caption{
    Relative error of the wavefunction value at the core
    radius for the hydrogen 2s orbital.
    Notations as in Figure~\ref{fig:1}.
  }
  \label{fig:4}
\end{figure}

\begin{figure}
  \centering
  \includegraphics{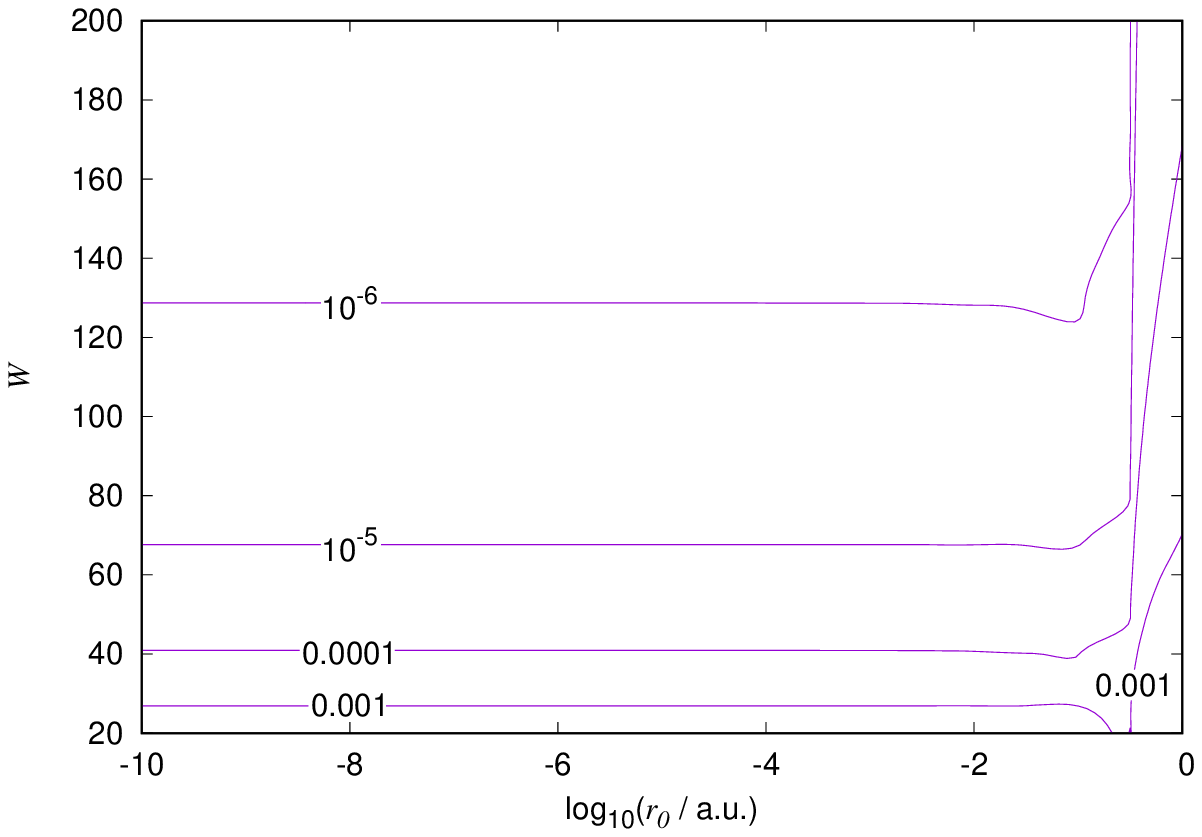}
  \caption{Hydrogen 2p orbital eigenenergy relative error.
    Notations as in Figure~\ref{fig:1}.
  }
  \label{fig:5}
\end{figure}

\begin{figure}
  \centering
  \includegraphics{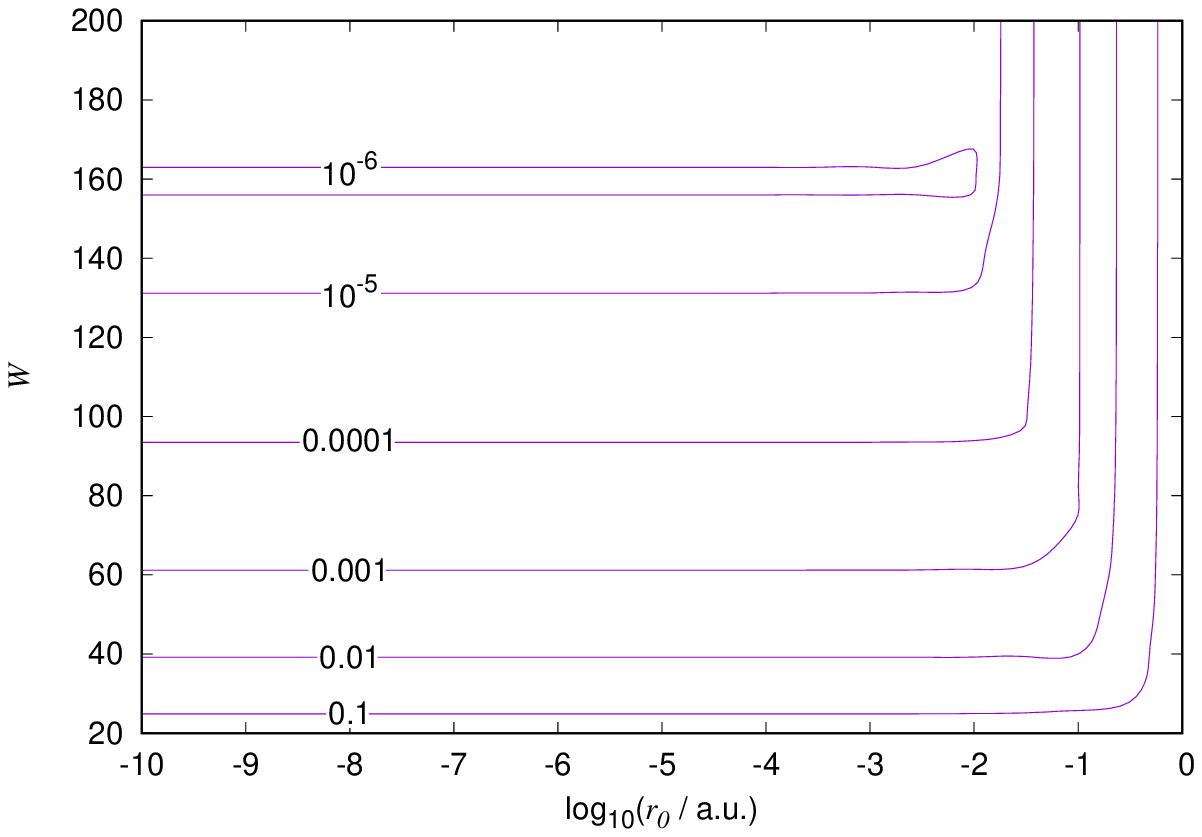}
  \caption{
    Relative error of the wavefunction value at the core
    radius for the hydrogen 2p orbital.
    Notations as in Figure~\ref{fig:1}.
  }
  \label{fig:6}
\end{figure}

\begin{figure}
  \centering
  \includegraphics{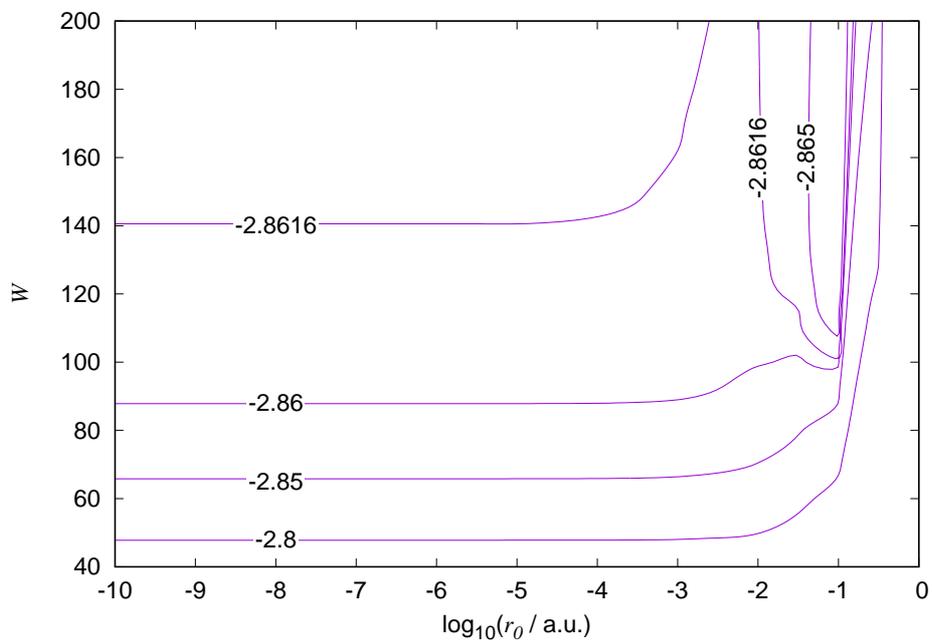}
  \caption{The ground state energy of helium atom. The HF limit
    is given in Table~\ref{tab:results}.
    Notations as in Figure~\ref{fig:1}.
  }
  \label{fig:7}
\end{figure}

\begin{figure}
  \centering
  \includegraphics{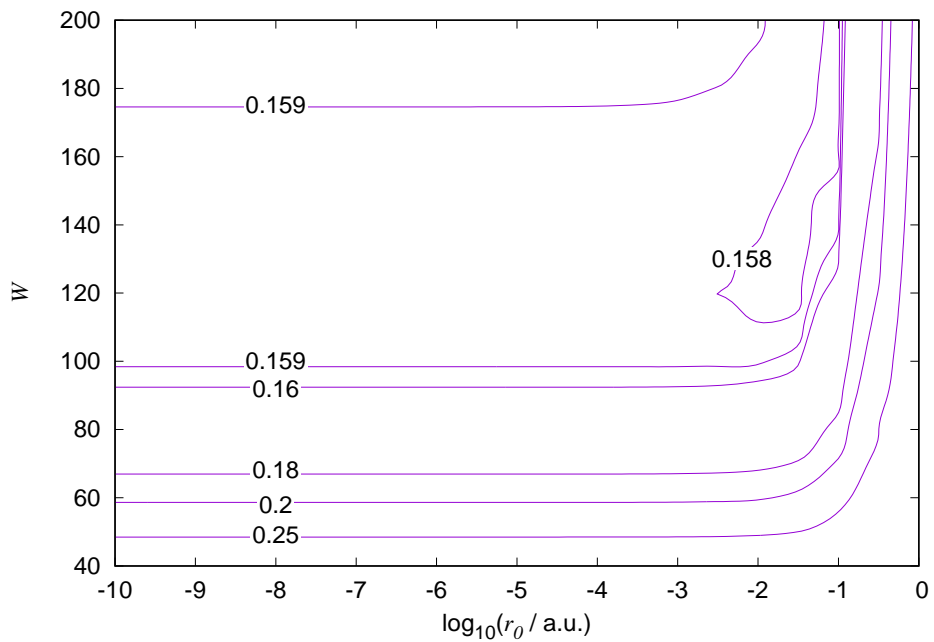}
  \caption{Relative error of the wavefunction value at the core
    radius for the 1s orbital of the ground state of the helium atom.
    Notations as in Figure~\ref{fig:1}.
  }
  \label{fig:8}
\end{figure}

\begin{figure}
  \centering
  \includegraphics{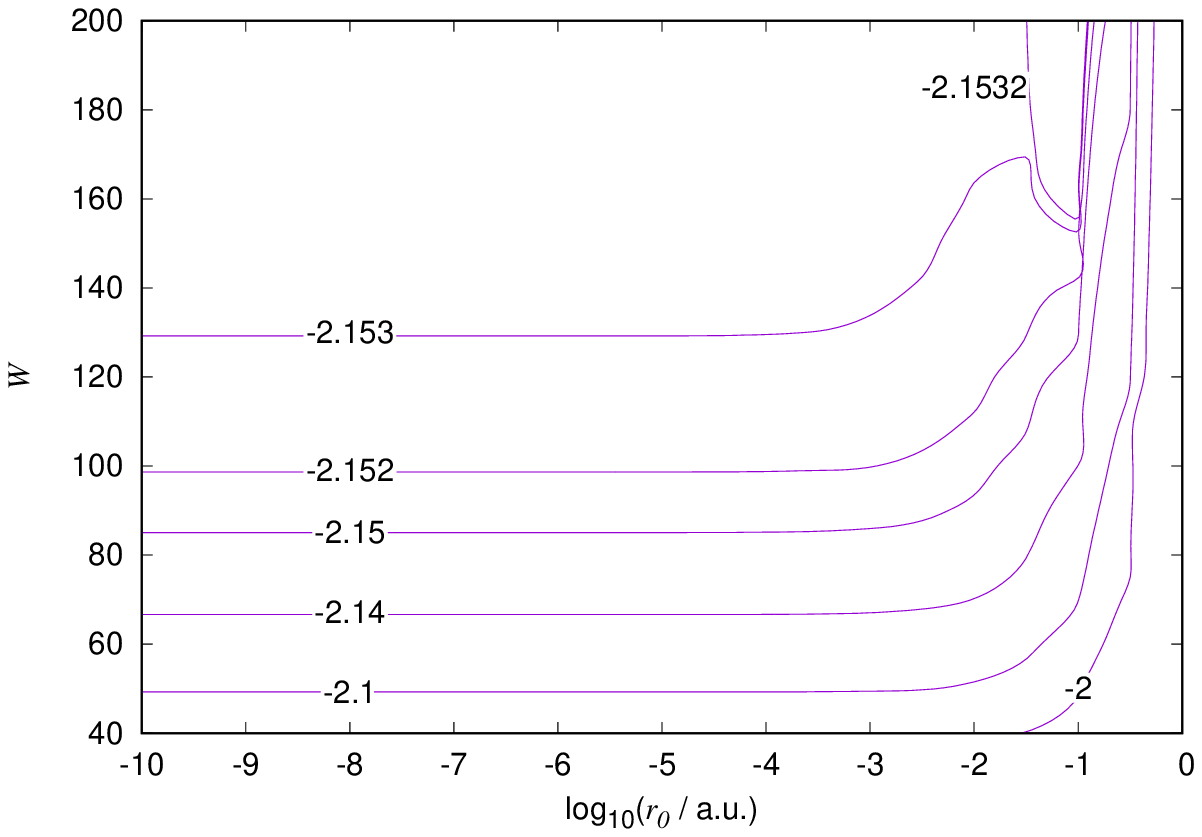}
  \caption{Helium 1s2s ${}^1\mathrm{S}$ total energy.
    The HF limit is given in Table~\ref{tab:results}.
    Notations as in Figure~\ref{fig:1}.
  }
  \label{fig:9}
\end{figure}

\begin{figure}
  \centering
  \includegraphics{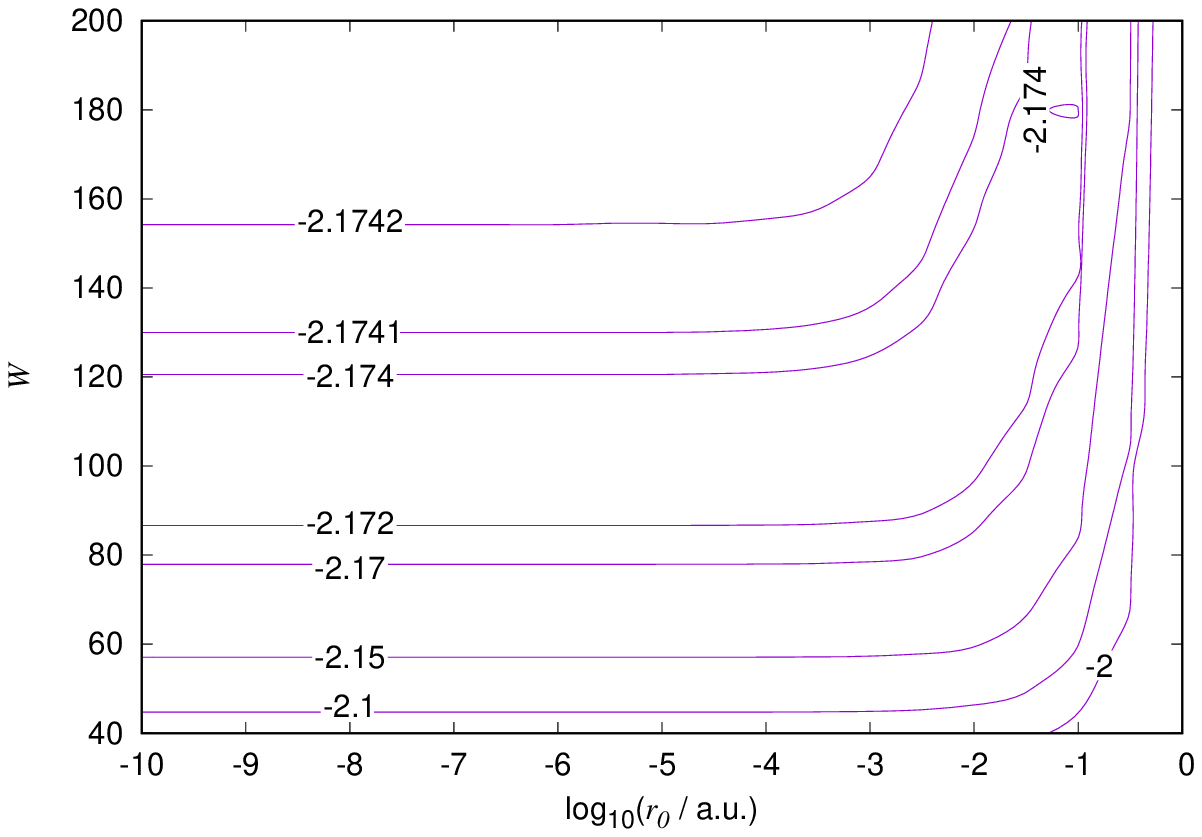}
  \caption{Helium 1s2s ${}^3\mathrm{S}$ total energy.
    The HF limit is given in Table~\ref{tab:results}.
    Notations as in Figure~\ref{fig:1}.
  }
  \label{fig:10}
\end{figure}

The results for the ground state of the hydrogen atom are presented in
figures \ref{fig:1} and \ref{fig:2}, for the 2s state in figures
\ref{fig:3} and \ref{fig:4}, and for the 2p state in figures
\ref{fig:5} and \ref{fig:6}.
%%It can be seen that the hydrogenic
%%orbitals are approximately continuous at \(r_0\).
The results of the
ground state of the helium atom are presented in figures \ref{fig:7}
and \ref{fig:8}.  The results for $\mathrm{He\;1s2s\;{}^1S}$ are given
in Figure~\ref{fig:9} and the results for $\mathrm{He\;1s2s\;{}^3S}$
in Figure~\ref{fig:10}.  As expected, the energy results are best for
large values of \(W\) and small values of \(r_0\).  Using 200 basis
functions for the helium ground state and computing the atom energies
for \(r_0 = 10^k\), \(k=-10,...,-1\) shows that atom energies are
equal up to seven decimals for \(r_0 \leq 10^{-6}\).  Similar
computation for hydrogen 1s orbital shows that the H 1s energy is
equal to \(-0.5 \; \mathrm{Ha}\) up to seven decimals for
\(r_0 \leq 0.01\).  For hydrogen 2s and 2p the corresponding limit is
\(r_0 \leq 0.01\), too.  We also found that when the number if basis
functions is sufficiently large for a given system there is an
approximate threshold value so that reducing
\(r_0\) below it does not make the accuracy of the computed energy better.
When the number of basis functions is sufficiently large and \(r_0\)
is sufficiently small the hydrogenic orbitals are approximately
continuous at \(r_0\).

The most accurate computations are in the upper left corners of the
figures. The orbitals of He 1s2s, except He 1s2s $^3\mathrm{S}$ 1s,
are not continuous at all at $r_0$
and no continuity plots are presented for them.
The computation results and exact results are given in Table~\ref{tab:results}.
For EPP--Galerkin method
the best energies (largest basis and smallest $r_0$) of the computed
systems are presented.
The OIW--Galerkin results with same number of basis functions and grid
spacing the same order of magnitude
as for the most accurate EPP results are given, too.
The accuracies of both of the methods depend on the grid
spacing.
The EPP--Galerkin method gives better results with the same number of
basis functions and larger grid spacing.
Results of Finite Difference Method are also given.
Note that for He 1s2s systems the OIW--Galerkin method with a basis set of 601
functions and finest grid point distance \(2^{-9} \;\mathrm{a.u.}\)
gives $E_{\mathrm{He\;1s2s\;{}^1S}} = -2.153\;\mathrm{Ha}$ and
$E_{\mathrm{He\;1s2s\;{}^3S}} = -2.174\;\mathrm{Ha}$, which are
approximately same as the results of the EPP--Galerkin method.

\begin{table}
  \centering
  \begin{tabular}{lrrrrrr}
    System &
    \(E_{\mathrm{exact}}\) / Ha &
    \(E_{\mathrm{EPP}}\) / Ha & \(h_{\mathrm{EPP}}\) / a.u.
    & \(E_{\mathrm{OIW}}\) / Ha & \(E_{\mathrm{FDM}}\) / Ha & \(N_{\mathrm{FDM}}\) \\
    \hline
    H 1s & -0.5 & -0.500000 & 0.075 & -0.498752 & -0.498031 & 1001 \\
    H 2s & -0.125 & -0.125000 & 0.125 & -0.124837 & -0.124741 & 2001 \\
    H 2p & -0.125 & -0.125000 & 0.125 & -0.124998 & -0.124995 & 2001 \\
    He $\mathrm{1s^2}$ & -0.28616800 \cite{ff1977} & -2.861629 & 0.075 & -2.834868 & -2.839010 & 1001 \\
    He $\mathrm{1s2s\;{}^1S}$ & -2.147 \cite{he1s2sb} & -2.153148 & 0.1 & -2.133579 & -2.132008 & 1001
    \\
    He $\mathrm{1s2s\;{}^3S}$ & -2.171 \cite{he1s2sb} & -2.174230 &
    0.1 & -2.154536 & -2.155362 & 1001
  \end{tabular}
  \caption{Computation results and parameters.
    \(E_{\mathrm{exact}}\) is the exact energy for H and HF limit for He.
    \(E_{\mathrm{EPP}}\)
    is the energy given by the EPP--Galerkin-method,
    \(h_{\mathrm{EPP}}\) is the grid spacing in the EPP--Galerkin
    method, \(E_{\mathrm{OIW}}\) is the energy given by the
    OIW--Galerkin method, \(E_{\mathrm{FDM}}\) is the energy given by
    the Finite Difference Method, and \(N_{\mathrm{FDM}}\) is the
    number of grid points in the Finite Difference Method.  For EPP
    the most accurate results are given. For OIW computations the
    number of basis functions is 201 and the finest grid spacing
    \(0.00625\;\mathrm{a.u.}\).}
  \label{tab:results}
\end{table}

%% The Hartree--Fock energy of the ground state of the helium atom is
%% \(-2.8616800\;\mathrm{Ha}\) \cite{ff1977}.
%% Reference \cite{he1s2s}
%% gives Hartree--Fock energies \(-2.147\;\mathrm{Ha}\) and
%% \(-2.171\;\mathrm{Ha}\) for
%% the $\mathrm{He\;1s2s\;{}^1S}$ and $\mathrm{He\;1s2s\;{}^3S}$
%% configurations, respectively.
%% Reference \cite{he2} gives quantities
%% \(E(\mathrm{He\;1s2s\;{}^1S}) - E(\mathrm{He\;1s^2}) =
%%   0.758\;\mathrm{Ha}\) and
%% \(E(\mathrm{He\;1s2s\;{}^3S}) - E(\mathrm{He\;1s^2}) =
%% 0.728\;\mathrm{Ha}\)
%% whence
%% \(E(\mathrm{He\;1s2s\;{}^1S}) = -2.104 \;\mathrm{Ha}\)
%% and
%% \(E(\mathrm{He\;1s2s\;{}^3S}) = -2.133 \;\mathrm{Ha}\).

Some of the computations using the diagonalization of the Hamiltonian operator
yield an unphysical state for the minimum eigenvalue.  For 1s and 2s
orbitals this eigenvalue seems to be about
\(-Z/r_0\) (in atomic units)
and the corresponding eigenvector
\(v(\wfb) = (-\delta_{k,0})_{k \geq 0}\). For hydrogen 2p orbital the
unphysical eigenvector does not appear.
The unphysical state remains the same during HF iteration of
$\mathrm{He\;1s^2}$,
$\mathrm{He\;1s2s\;{}^1 S}$, and $\mathrm{He\;1s2s\;{}^3 S}$.
The physical admissibility of the
wavefunctions \(P_{nl}(r)\) was characterized by condition
\begin{equation}
  \lim_{r \to 0} P_{nl}(r) = 0 .
\end{equation}
We checked this condition by extrapolating solutions \(P_{nl}(r)\)
polynomially at \(r = 0\). Actually we extrapolate polynomially
\(\wfb(s)\) at \(s=-r_0\) using some points \(s\) near 0.
Note that Fischer and Defranceschi
\cite{fd1998} also get unphysical states in wavelet
computations of hydrogen-like atoms.
Their iteration scheme yields an unphysical result that is actually
the mathematical ground state corresponding to the pseudopotential.

\section{Conclusions}

%% There were six correct decimals in the most accurate energy of
%% hydrogen 1s, 2s, and 2p orbitals.
%% Helium 1s2s computations gave four correct decimals in the
%% most accurate computations.
EPP--Galerkin method gives seven correct decimals for the hydrogenic 1s
orbital, six correct decimals for the hydrogenic 2s and 2p orbitals, and
four correct decimals for He \(\mathrm{1s}^2\).
For He 1s2s \({}^1\mathrm{S}\) and \({}^3\mathrm{S}\) we get energies
close to the HF limit.
OIW--Galerkin method with finest grid spacing
\(0.0625\;\mathrm{a.u.}\) gives energies with two to five correct
decimals.  The grid size of OIW--Galerkin calculations is smaller
compared to the EPP--Galerkin calculations.  Finite Difference Method
yields rather inaccurate results even though the grid spacing is
considerably smaller compared to the EPP--Galerkin calculations.
%% Computation of hydrogen 1s orbital in
%% \cite{hrr2004} gave relative energy error about $10^{-5}$ with 200
%% basis functions and EPP gives about $10^{-6}$ with the same number of
%% basis functions. Hydrogen 2s computations in \cite{hrr2004} gave
%% relative energy error about $10^{-6}$ and EPP gives similar error with
%% 160 basis functions. The smallest relative energy error of the helium
%% ground state computed with EPP was about $0.00002$ (200 basis
%% functions) whereas in \cite{hrr2004} it was about $0.00004$ (180 basis
%% functions).
We were able to get results near the Hartree--Fock limit by using
large enough basis and small enough parameter \(r_0\).
It turned out that EPP--Galerkin method yields better
methods than the OIW--Galerkin method and
considerably better results than the Finite Difference Method.

\printbibliography

\end{document}